\documentclass[final, 5p, times]{elsarticle}
\DeclareUnicodeCharacter{2113}{\ensuremath{\ell}}
\usepackage{subcaption}
\usepackage{ulem}
\usepackage{xcolor}
\usepackage{amsmath}
\usepackage{etoolbox}
\usepackage[left]{lineno}
\makeatletter
\def\@alph#1{\number#1}
\def\ps@pprintTitle{%
 \let\@oddhead\@empty
 \let\@evenhead\@empty
 \let\@oddfoot\@empty
 \let\@evenfoot\@empty
}
\makeatother
\begin{document}
\sloppy
\begin{frontmatter}

\title{Characterization of a CsI(Tl) Scintillator Coupled to a SiPM at Cryogenic Temperatures}

\author[inst1]{M. Mirzakhani}
\author[inst1]{R. Mahapatra}
\author[inst1]{M. Platt}

\affiliation[inst1]{organization={Department of Physics and Astronomy, Texas A\&M University },
            addressline={578 University Dr}, 
            city={College Station},
            postcode={77840}, 
            state={TX},
            country={US}}

\begin{abstract}
We report on the scintillation characterization of a \(1\,\text{cm}^3\) CsI(Tl) crystal coupled to a \(6\times6\,\text{mm}^2\) SiPM read out with a custom transimpedance amplifier at cryogenic temperatures. The crystal was prepared with optical-grade surfaces and enclosed in a multi-layer shielding system to suppress both ambient light and background radiation. The detector response was systematically studied at temperatures ranging from 100\,K to room temperature. A figure of merit, defined as the ratio of the 59.5\,keV peak position to the baseline resolution, revealed a pronounced maximum at 175\,K, indicating optimal photon detection efficiency at this operating point. Pulse shape analysis further demonstrated clear temperature-dependent variations in scintillation decay characteristics. Using data collected at 175\,K, a baseline resolution calculated \(\sigma \approx 1~\text{keV}\). This corresponds to an effective threshold of \(3\sigma \approx 3~\text{keV}\), confirming the capability of the CsI(Tl)–SiPM system to operate at low thresholds, underscoring its potential for rare-event searches and low-energy particle detection.
\end{abstract}

\end{frontmatter}

\section{Introduction}

Thallium-doped cesium iodide (CsI(Tl)) is a well-established inorganic scintillator widely used in nuclear and particle physics due to its high light yield, relatively fast response, and compatibility with a variety of photodetectors. Its emission peak around 550\,nm aligns well with the spectral sensitivity of silicon photomultipliers (SiPMs), making it a particularly attractive choice for compact, high-efficiency scintillation detectors. While traditionally read out using photomultiplier tubes (PMTs), recent advances in SiPM technology have enabled the use of CsI(Tl) in low-voltage, cryogenic, and magnetic field-sensitive environments.

CsI(Tl) crystals have demonstrated robust performance in applications ranging from gamma spectroscopy to calorimetry and low-background experiments. At room temperature, CsI(Tl) exhibits a light yield exceeding 60,000 \(\frac{\mathrm{photons}}{\mathrm{MeV}}\)
~\cite{Mikhailik_2015, cryst12111517, MIANOWSKA2022166600}, and its scintillation decay time, typically on the order of 1\,\textmu s, supports pulse shape discrimination (PSD) under appropriate timing conditions. When operated at cryogenic temperatures, CsI(Tl) maintains stable light output and benefits from reduced thermal noise in SiPMs, enhancing signal-to-noise ratios in precision measurements.

Recent studies have compared CsI(Tl) and undoped CsI in the context of particle identification. For example, Longo et al.~\cite{Longo_2022} demonstrated that while undoped CsI exhibits faster decay components suitable for neutron-gamma discrimination, CsI(Tl) still offers meaningful PSD capabilities, particularly when coupled to fast digitizers and optimized readout electronics. Furthermore, the COHERENT collaboration has deployed CsI(Tl) detectors for coherent elastic neutrino-nucleus scattering (CEvNS) measurements, demonstrating their sensitivity to low-energy nuclear recoils~\cite{Yang}.

In addition to fundamental research, CsI(Tl) coupled to SiPMs holds promise for dark matter searches, where low thresholds and background suppression are paramount. While undoped CsI may offer advantages in timing resolution and pulse shape contrast, CsI(Tl) provides superior light yield and more consistent optical properties, especially when combined with SiPM arrays that match its emission spectrum~\cite{kim2024scintillationcharacteristicsundopedcsi}. This combination enables scalable, high-granularity detector designs with minimal dead space and improved energy resolution.

In this work, we report on the characterization of a CsI(Tl) crystal coupled to SiPMs over a range of temperatures. We present measurements of light yield, decay time, and pulse shape behavior under gamma-ray excitation, with an emphasis on evaluating the detector’s suitability for rare-event searches and low-background experiments. These results offer valuable insight into the performance of CsI(Tl)-SiPM systems and their potential for future cryogenic applications.

\section{Experimental Setup}

\subsection{Detector Assembly and Readout}

In this study, we characterized scintillation light yield at cryogenic temperatures using a \(1\,\text{cm}^3\) CsI(Tl) crystal optically coupled to a \(6\times6\,\text{mm}^2\) silicon photomultiplier (SiPM). The crystal was cut into a cubic shape with a precision wire saw and polished sequentially with fine sandpapers and liquid agents to achieve optical-grade surfaces. To optimize light collection, different wrapping materials were tested. In one configuration, all crystal faces were covered with Teflon for internal reflection and an outer layer of aluminum foil as an external reflector. While this combination is widely used at room temperature, it is less suitable at cryogenic conditions due to thermal conductivity concerns. Therefore, in our low-temperature setup, silver tape was chosen as the primary wrapping material to ensure both reflectivity and better thermal compatibility. One face of the crystal, corresponding to the photodetector coupling side, was partially unwrapped by removing a \(6 \times 6\,\text{mm}^2\) section, allowing direct optical contact with the SiPM.

The SiPM used in this setup was Microfc-60035-SMT-TR1 from ONSEMI, selected for its compact active area and cryo-compatibility. To enhance the signal-to-noise ratio and ensure stable readout at low temperatures, the SiPM was connected to a custom amplifier circuit. Optical grease was applied between the crystal and the SiPM to minimize photon loss at the interface and improve optical coupling.

The complete detector assembly was mounted inside a CTI-Cryogenics Cryodyne Refrigeration System (Model 350), which provides a base temperature of approximately 14\,K. Bias voltages of 34\,V and 5\,V were applied to the SiPM and amplifier circuit, respectively. The SiPM featured dual-channel readout—fast and slow timing outputs. In the primary phase of the experiment, the slow timing output was used and connected to a PicoScope 5442D digitizer via SMA coaxial cables. This digitizer has a 250\,MHz analog bandwidth, suitable for resolving the scintillation pulse shapes.

\begin{figure}[htbp]
  \centering
    \includegraphics[height=4.8cm, width=\linewidth]{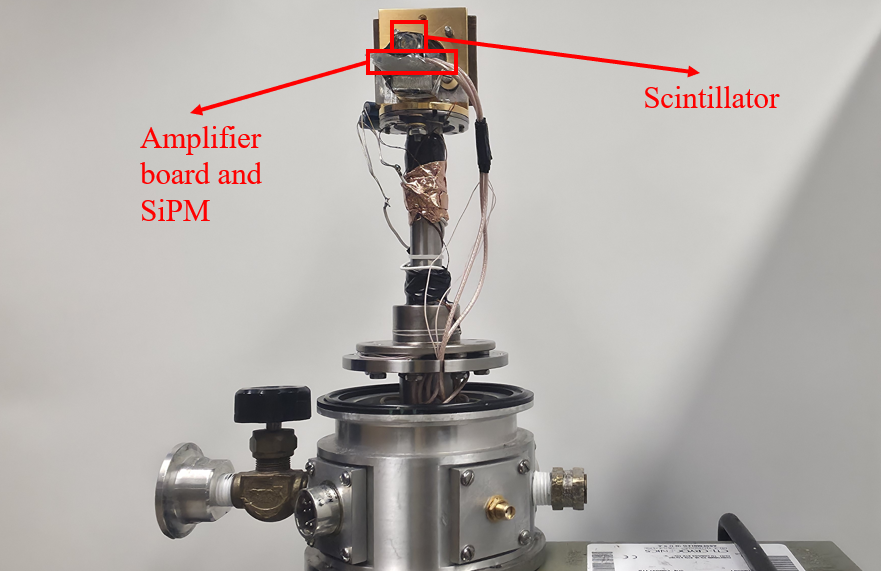}
   \caption{Actual picture of detector and scintillator crystal.}
  \label{fig:sample}
\end{figure}

To ensure good thermal contact while maintaining electrical insulation, the rear of the amplifier board was coated with a thin layer of hot glue. This provided insulation from the copper mounting tape used to thermally anchor the assembly. The crystal and SiPM module were then mechanically secured and thermally coupled using copper tape, as shown in Fig.~\ref{fig:sample} (right).

\subsection{Stability of Cryostat}

Before commencing any data acquisition, it is critical to ensure the thermal stability of the cryostat at the target temperatures. For this purpose, a calibrated silicon diode thermometer (Lake Shore DT-670) was used to monitor the temperature inside the cryogenic chamber. The sensor was interfaced with a Lake Shore Model 330 temperature controller, which was preloaded with a custom calibration curve to ensure accurate temperature readings.

The calibration data for the DT-670 sensor were imported into the controller by connecting it to a PC via a GPIB interface. Using National Instruments Measurement \& Automation Explorer (NI MAX), the calibration points were uploaded and verified. This procedure ensured that the temperature readings corresponded precisely to the diode’s characteristics across the desired temperature range.

To achieve stable temperature control at various operating points, two Lake Shore HTR-25-100 resistive heaters (\(25\,\Omega\), \(100\,\text{W}\)) were mounted in close proximity to the sample holder. These heaters provided localized heating power to counteract thermal gradients and maintain equilibrium conditions. The cryostat stability was then evaluated by setting different temperature points and monitoring their response over time, ensuring that the operating temperature remained constant without significant fluctuations during measurements.

\subsection{Shielding}

\begin{figure}[htbp]
  \centering
    \includegraphics[height = 6.2 cm, width=\linewidth]{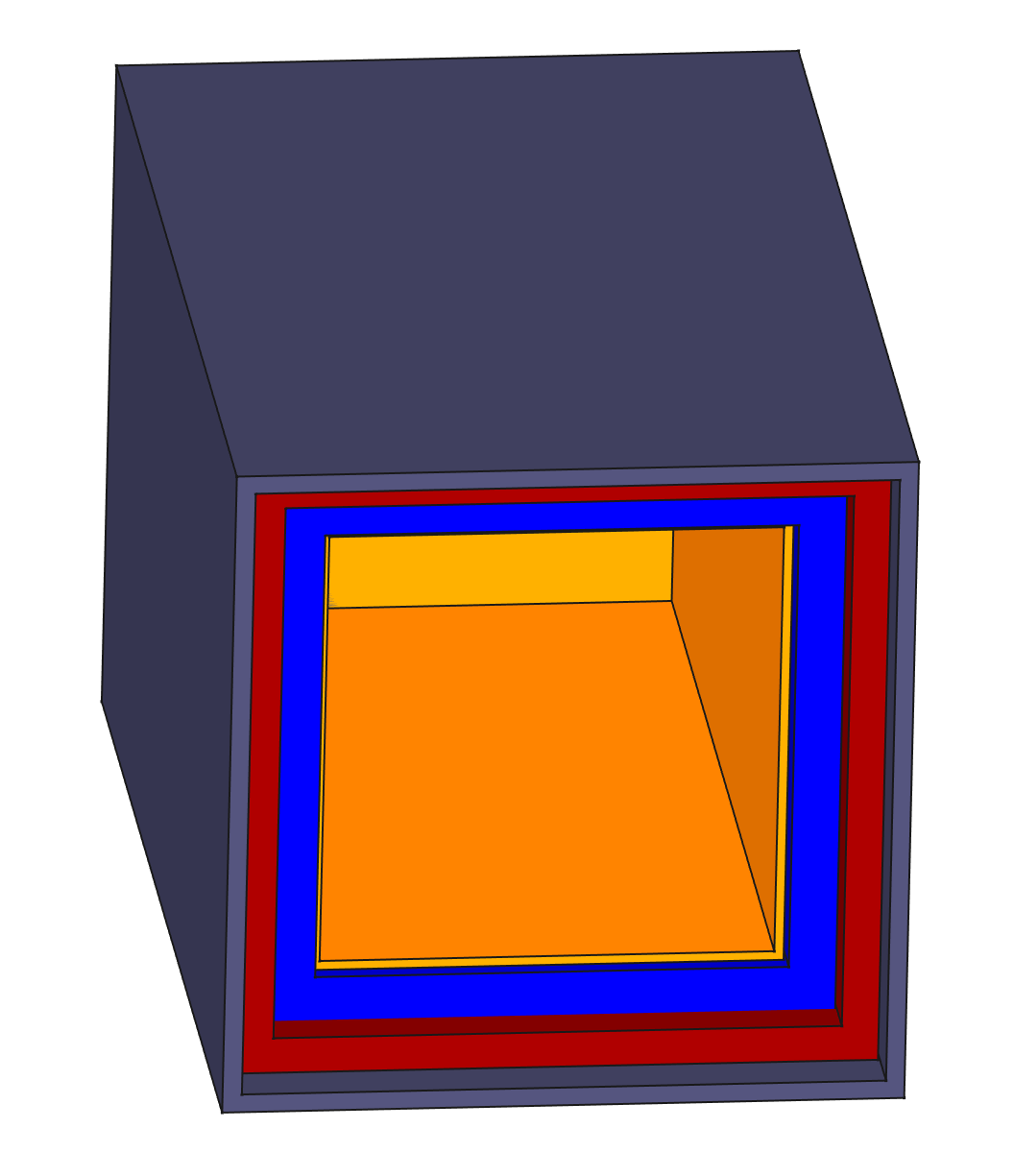}
   \caption{3D schematic of the cryostat shielding configuration designed in AutoCAD, showing the layer sequence from inner to outer: a \(0.2\,\text{mm}\) copper layer, a \(5\,\text{mm}\) PLA filament, a \(4\,\text{mm}\) lead sheet, and a \(2\,\text{mm}\) plastic rubber enclosure.}
  \label{fig:Schematic Shielding}
\end{figure}

To minimize stray ambient light and suppress background radiation, a multi-layer shielding system was employed around the cryostat chamber. The innermost layer consisted of a copper tape to cover cryostat windows, custom-designed PLA filament enclosure, fabricated using a 3D printer to ensure a precise fit. This layer was fully covered with \(4\,\text{mm}\)-thick lead sheets to attenuate external gamma radiation. Finally, an additional outer layer of blackout material was applied to further reduce residual light leakage. A schematic of the shielding configuration is shown in Fig.~\ref{fig:Schematic Shielding}.

Since a significant fraction of background radiation originates from the ground, an extra \(4\,\text{mm}\) lead plate was placed underneath the cryostat to suppress upward-directed events. This layered approach provided both optical isolation and enhanced radiation shielding, ensuring a low-background environment for scintillation measurements.

\subsection{Stability of SiPM}

To accurately characterize the CsI(Tl) scintillator at various temperatures, it was essential to verify that the SiPM operated reliably across the full temperature range of the experiment. For this purpose, the SiPM was biased and gradually cooled in a completely dark environment. During the cooling process, data were continuously recorded from the SiPM’s slow-time readout channel until the system reached its base temperature.

\begin{figure}[htbp]
  \centering
    \includegraphics[width=\linewidth]{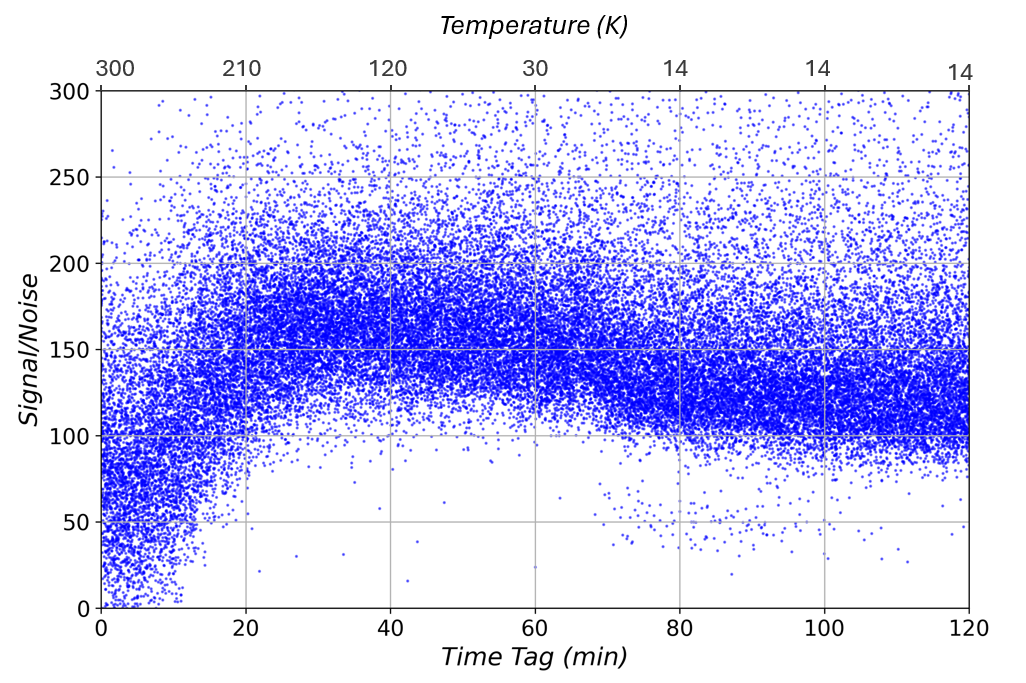}
   \caption{Signal to Noise Ratio (SNR) vs. time tag of each event (triggered moment) until cooling down to 14K.}
  \label{fig:S over N vs Time}
\end{figure}

As expected, lowering the temperature suppressed thermal noise, resulting in an improved signal-to-noise ratio (SNR). This improvement arises from two main sources of noise reduction: (1) a decrease in thermal excitations within the SiPM caused by random motion of charge carriers, and (2) reduced contributions from the intrinsic background of the scintillator. A plot of signal over noise as a function of time during cool-down (equivalently temperature) confirms this trend, showing a significant increase in SNR approximately 20 minutes after the start of cooling, corresponding to a temperature of around 210\,K. For this purpose, the signal was defined as the integrated area under the raw pulse, while the noise was calculated as the sum of the pre-pulse sample points. As shown in Fig.~\ref{fig:S over N vs Time}, confirming the robust operation of the SiPM across the studied temperature range.

Due to carrier freeze-out in the silicon lattice, a significant reduction in SiPM gain is expected below 100\,K. This behavior is evident in Fig.~\ref{fig:S over N vs Time}, where a noticeable drop occurs after approximately 60 minutes, corresponding to 100\,K. However, because the SiPM material is intentionally doped, complete freeze-out is avoided, allowing the device to remain operational at temperatures well below this threshold. As a result, only a modest decrease in gain is observed rather than a complete loss of functionality.

\subsection{Amplifier Board}

\begin{figure}[htbp]
  \centering
  \begin{subfigure}[t]{0.24\textwidth}
    \centering
    \includegraphics[height=4.8cm, width=\linewidth]{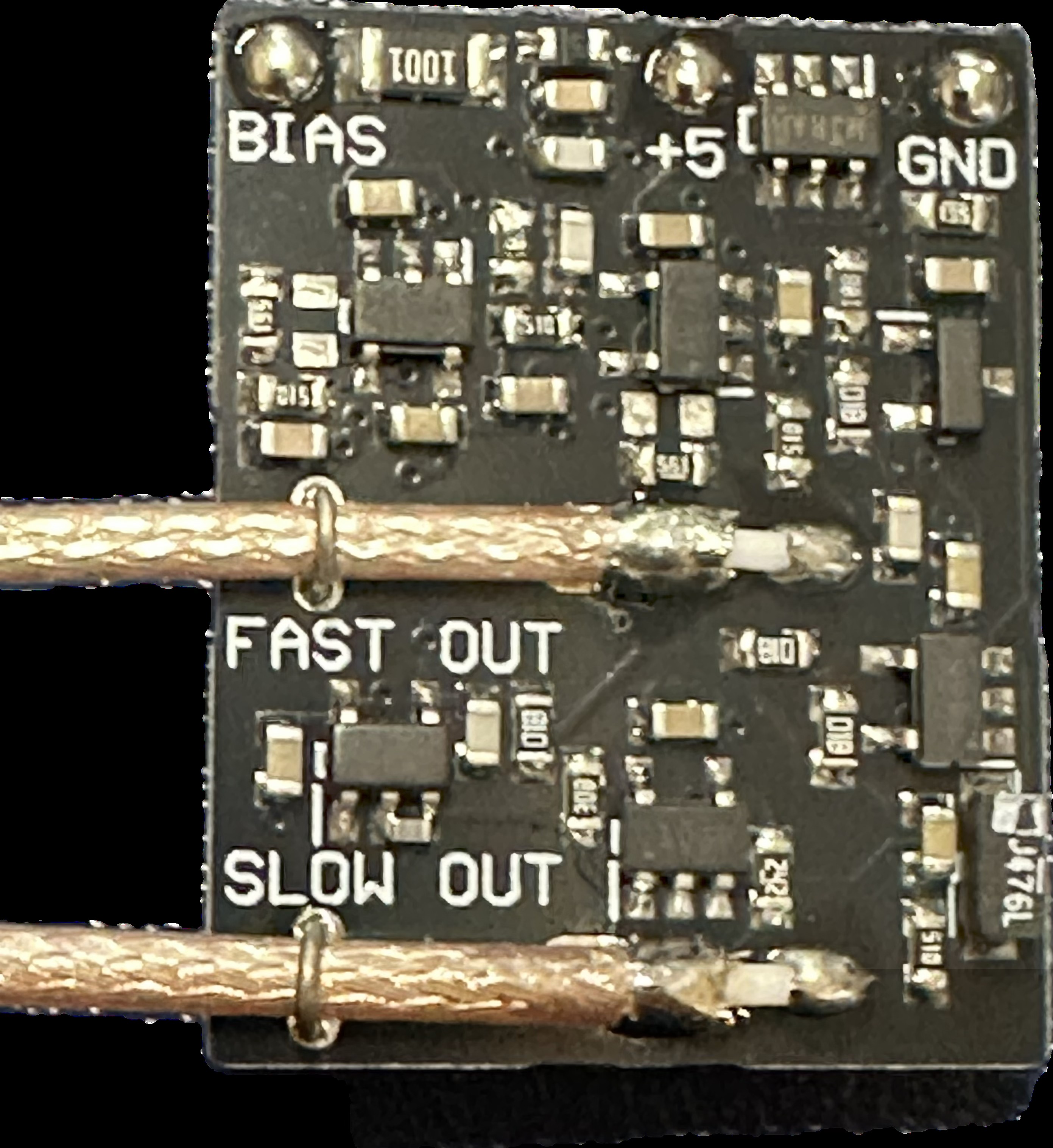}
    \caption{}
  \end{subfigure}\hfill
  \begin{subfigure}[t]{0.24\textwidth}
    \centering
    \includegraphics[height=4.8cm, width=\linewidth]{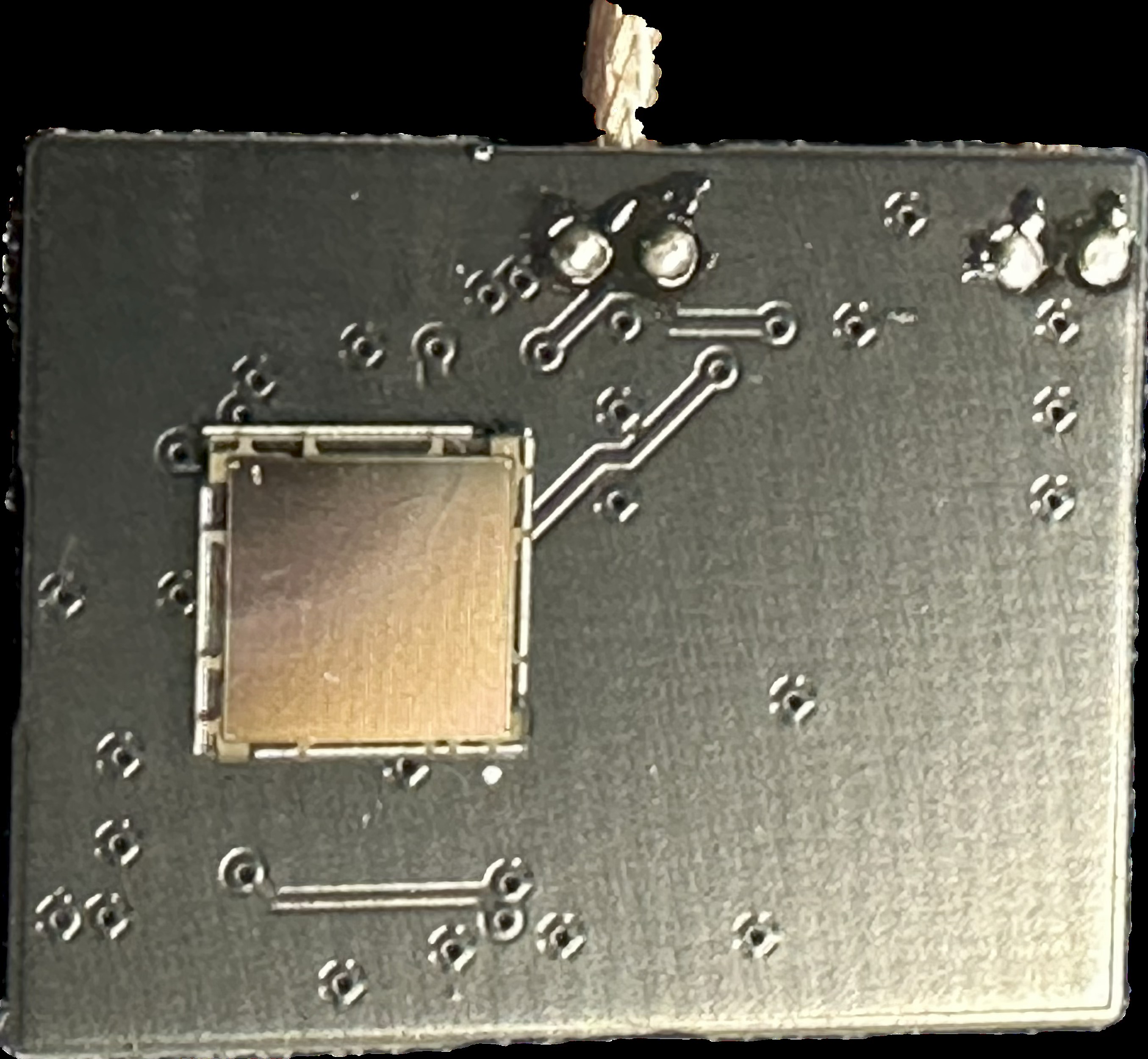}
    \caption{}
  \end{subfigure}

  \caption{(1) Actual picture of amplifier circuit, (2) the attached SiPM to rear side of amplifier board.}
  \label{fig:board}
\end{figure}

To clarify the readout chain used for SiPM signal amplification, we describe here the design and implementation of the amplification circuit. A photograph of the assembled circuit board, with the SiPM mounted on its reverse side, is shown in Fig.~\ref{fig:board}.

\begin{figure*}[htbp]
  \centering
  \includegraphics[width=\textwidth]{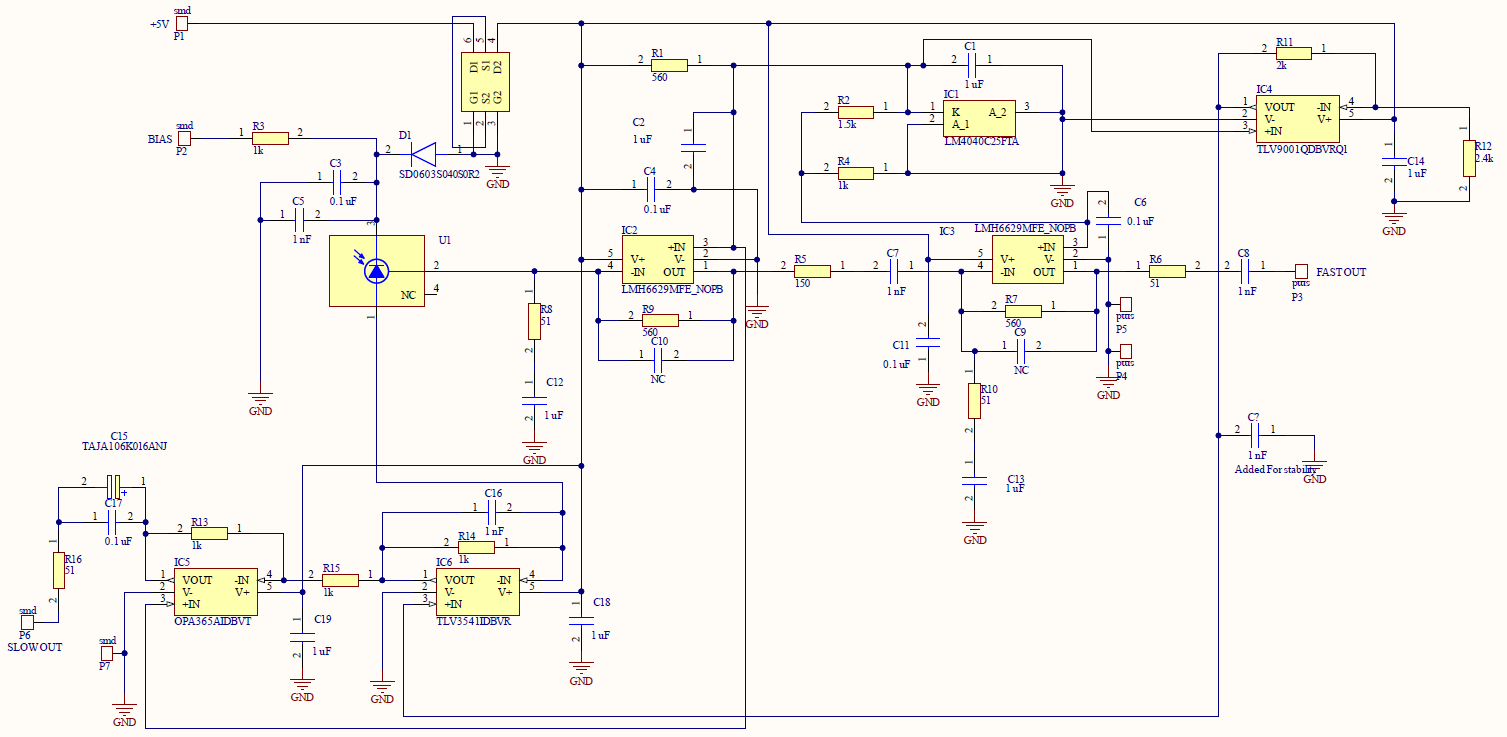}
  \caption{ Custom amplifier board designed for this project in collaboration with Atom Spectra Australia.
  }
  \label{fig:Amplifier}
\end{figure*}

A custom SiPM readout circuit was designed to achieve efficient signal amplification and bandwidth suitable for photon detection. As illustrated in Fig.~\ref{fig:Amplifier}, the SiPM photocurrent is fed into a two-stage front-end. The initial stage utilizes a transimpedance amplifier, configured with precision low-noise operational amplifiers (e.g., LMH6629), to convert the fast and low-level SiPM current pulses into voltage. Further active stages (including IC2, IC3, and IC5) provide additional shaping, buffering, and drive capability, with careful feedback and decoupling—via capacitor and resistor networks—to optimize stability and minimize noise. Outputs are routed via designated connectors for both Fast and Slow timing paths, enabling flexible downstream analysis. The layout includes multiple supply decoupling capacitors and biasing networks to ensure low-noise operation and robust performance for time-sensitive photon counting applications. We also have implemented reverse polarity protection to the amplifier board, current limiting resistor to SiPM bias to avoid SiPM burnout if biased under ambient light, Shottky diode for SiPM protection against forward biasing. This amplifier board was custom developed by Atom Spectra Australia in collaboration with our research group.

After verifying the essential steps required to establish a fully operational setup, the following section presents the experimental results. These include the determination of the optimal operating temperature for the SiPM, as well as the evaluation of the baseline resolution and energy threshold achieved in this study.

\section{Results}
\subsection{Light Output Threshold}

Following the installation of the CsI(Tl) crystal inside the cryostat, the SiPM was coupled to the unwrapped face of the crystal and properly biased. All measurements were carried out in a dark environment to suppress stray ambient light. Data acquisition was performed at various temperatures using a PicoScope 5442D digitizer with a sampling rate of \(10\,\text{MS/s}\).

\begin{figure}[htbp]
  \centering
    \includegraphics[width=\linewidth]{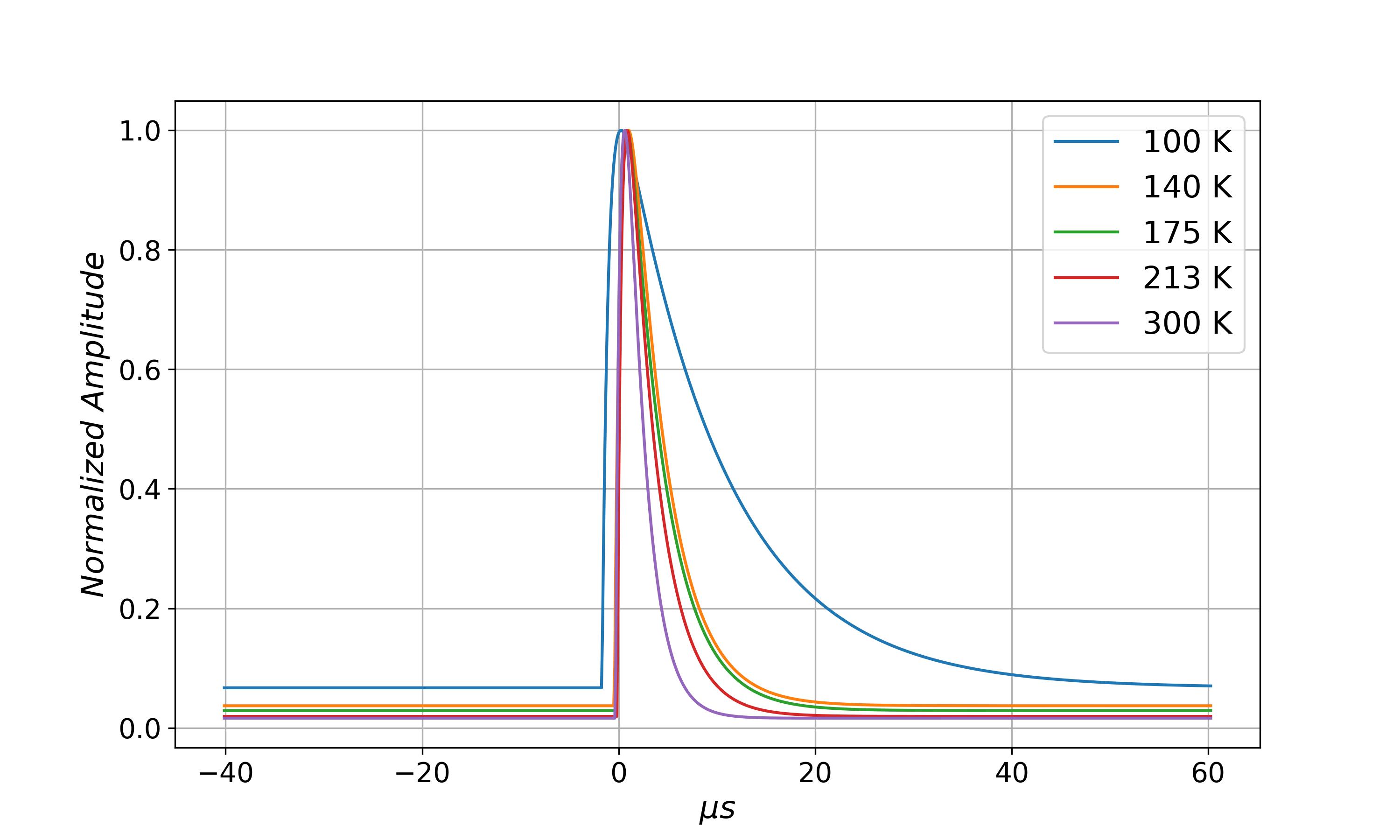}
   \caption{Pulse template fits at different temperatures. As the temperature decreases, the pulse decay time increases.}
  \label{fig:template_Fit_Am}
\end{figure}

The evolution of scintillation dynamics with temperature was further examined using representative pulse templates fitted at selected temperatures, as shown in Fig.~\ref{fig:template_Fit_Am}. These results clearly demonstrate variations in scintillation decay time and overall pulse morphology as the temperature decreases, reflecting the strong temperature dependence of carrier transport and recombination processes in CsI(Tl). At higher temperatures, charge trapping and recombination occur more rapidly due to enhanced carrier mobility and thermal excitation. As the temperature decreases, however, thermal release (untrapping) of electrons from \( \mathrm{Tl}^{0} \) centers becomes increasingly suppressed, slowing the recombination process and resulting in longer scintillation decay times~\cite{PhysRevApplied.7.014007, KIM2025103150}.

\begin{figure}[htbp]
  \centering
    \includegraphics[width=\linewidth]{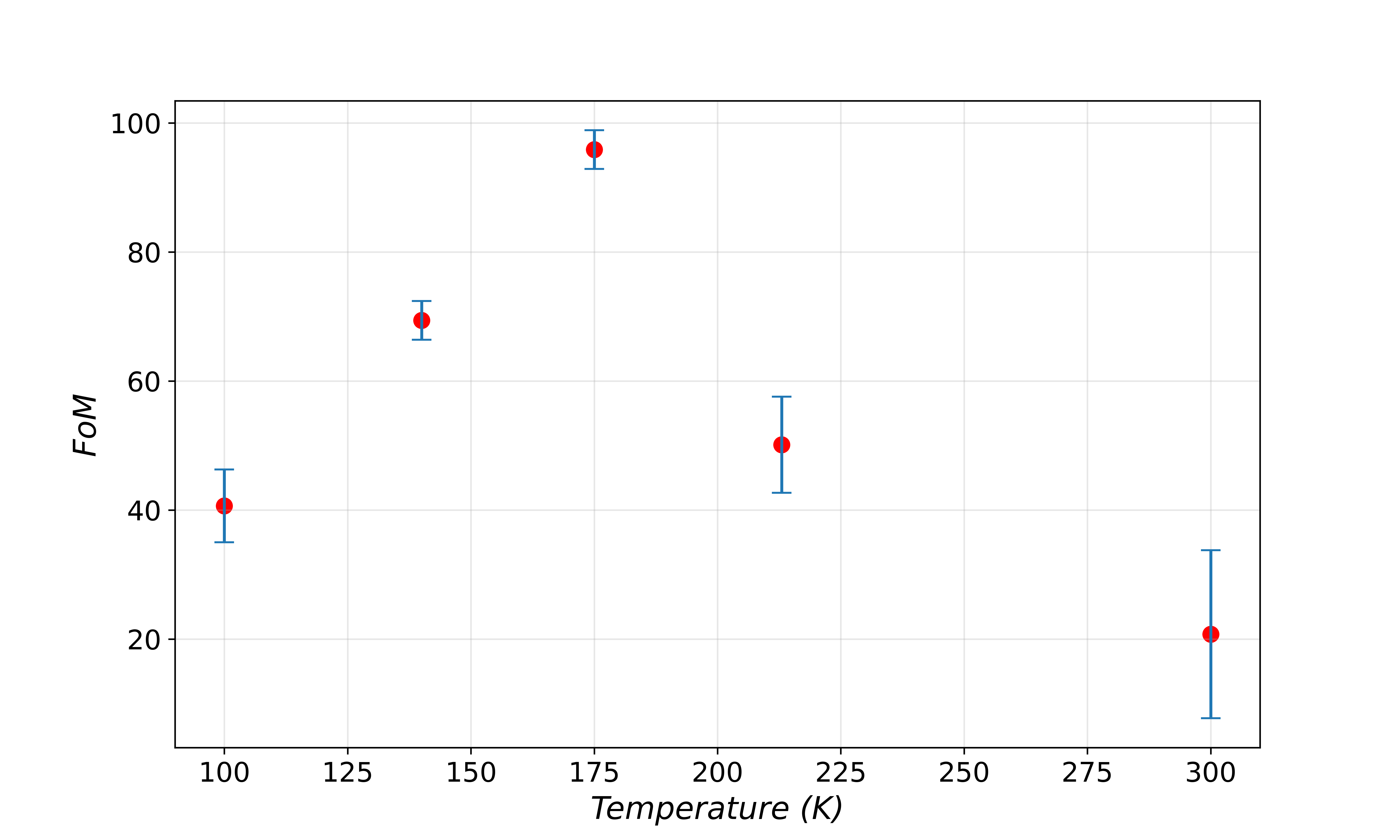}
   \caption{Detector Figure of Merit resulting division of peak position corresponding to 59.5\,keV to the baseline resolution at defined temperatures.}
  \label{fig:Figure of Merit}
\end{figure}

To determine the optimal operating temperature of the detector, spectra were generated by integrating pulse waveforms processed with the Optimal Filter (OF) method at each temperature. As a performance metric, we define a figure of merit (FoM) as the ratio of the 59.5\,keV peak position from the \textsuperscript{241}Am source to the corresponding baseline resolution at the same temperature, both expressed in arbitrary units. This FoM provides a direct measure of the performance of the detector. Representative results obtained with a proper hardware threshold are shown in Fig.~\ref{fig:Figure of Merit}. A pronounced maximum in FoM is observed near 175\,K, followed by a decline at lower temperatures. This trend indicates that the SiPM attains its highest photon detection efficiency in this range, leading to enhanced light collection. Data below 100\,K are not included, since—as discussed earlier—carrier freeze-out in the silicon lattice causes a substantial reduction in gain, resulting in even lower light output than at 100\,K.

After identifying the optimal operating temperature, the hardware threshold was carefully optimized. Preliminary scans across different settings showed that \(2\,\text{mV}\) provided the best compromise, effectively suppressing electronic noise while maximizing retention of genuine scintillation events, and yielding the lowest practical threshold for reliable signal discrimination.

With the operating conditions established, calibration was performed using \textsuperscript{241}Am and \textsuperscript{55}Fe radioactive sources, each with an activity of \(10\,\mu\text{Ci}\). The \textsuperscript{241}Am source provided well-defined emission lines at 26.5\,keV and 59.5\,keV, while the \textsuperscript{55}Fe source was employed both to probe the 5.9\,keV X-ray peak and to assess the effective detection threshold. Because CsI(Tl) exhibits a well-known non-proportional scintillation response, particularly at cryogenic temperatures where carrier transport and recombination dynamics are strongly energy dependent, a quadratic calibration function was adopted to more accurately relate the integrated pulse area to the deposited energy.  

Data collection was divided into three separate runs:  
1) To compensate for the short attenuation length of 5.9\,keV X-rays, the \textsuperscript{55}Fe source was placed directly on top of the crystal to maximize the event rate.  
2) The \textsuperscript{241}Am source was mounted on top of the sample, to make sure the condition is similar for both sources.  
3) A background run was acquired in random-trigger mode to characterize electronic noise and validate the effective detector threshold.  

\begin{figure}[h!]
  \centering
    \includegraphics[width=8 cm]{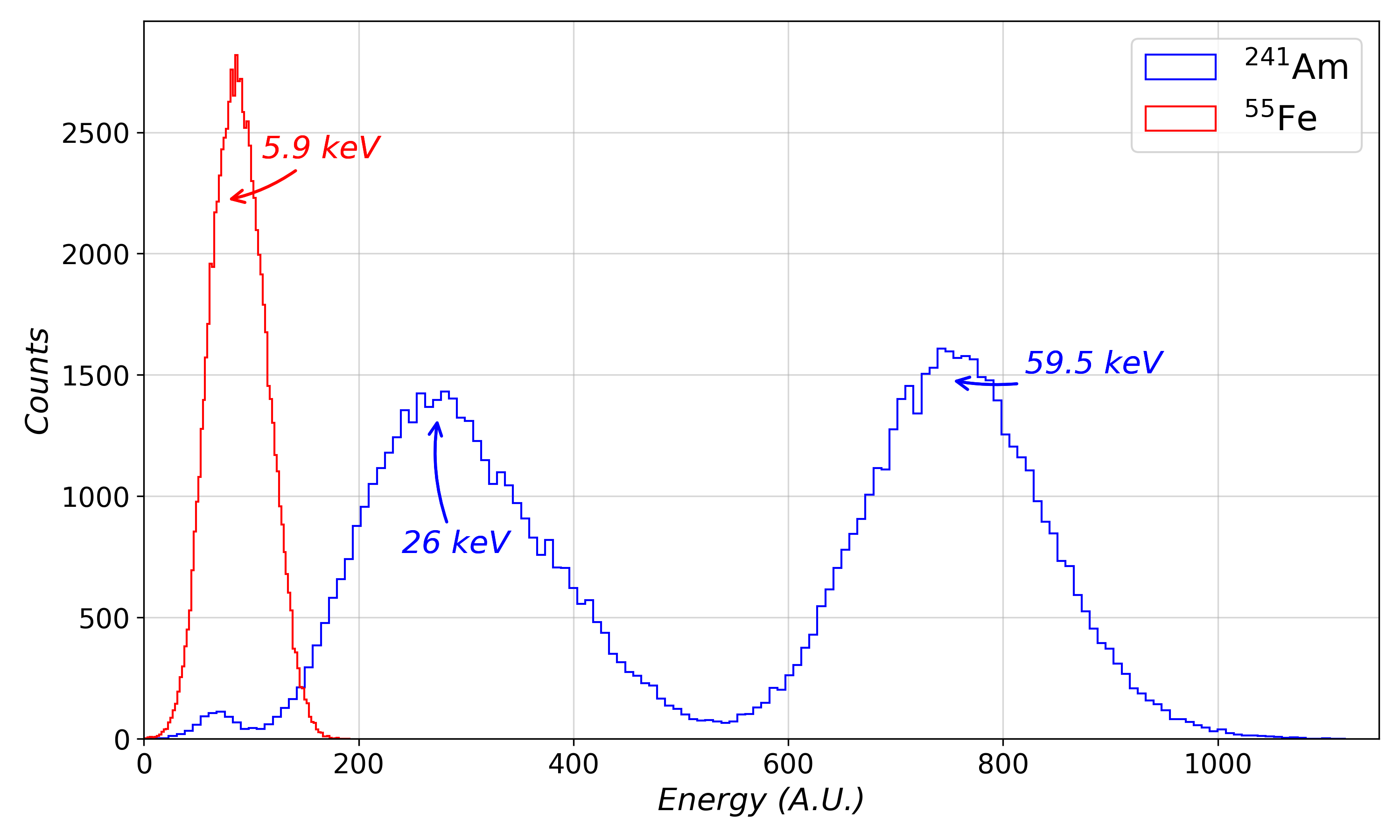}
   \caption{Light output spectrum the using area event integration.}
  \label{fig:spectrum}
\end{figure}

Each run lasted approximately six hours, and the combined spectrum from the two radioactive sources is shown in Fig.~\ref{fig:spectrum}. 

In this work, we applied a novel analysis technique to reject noise and improve the quality of the raw energy spectrum. The presence of noisy events is a common challenge in scintillation measurements and often complicates calibration. Conventional approaches, such as applying a \(\chi^{2}\) cut, are typically used to select signal-like events. In contrast, we employed Principal Component Analysis (PCA) \cite{Liu:2019jxg, Dong:2023nir, PhysRevC.103.034909} to identify and select high-quality events, particularly those corresponding to emission lines from radioactive calibration sources.  

\begin{figure}[htbp]
  \centering

      \begin{subfigure}[t]{0.55\textwidth}
        \centering
        \includegraphics[height=4.8cm, width=\linewidth]{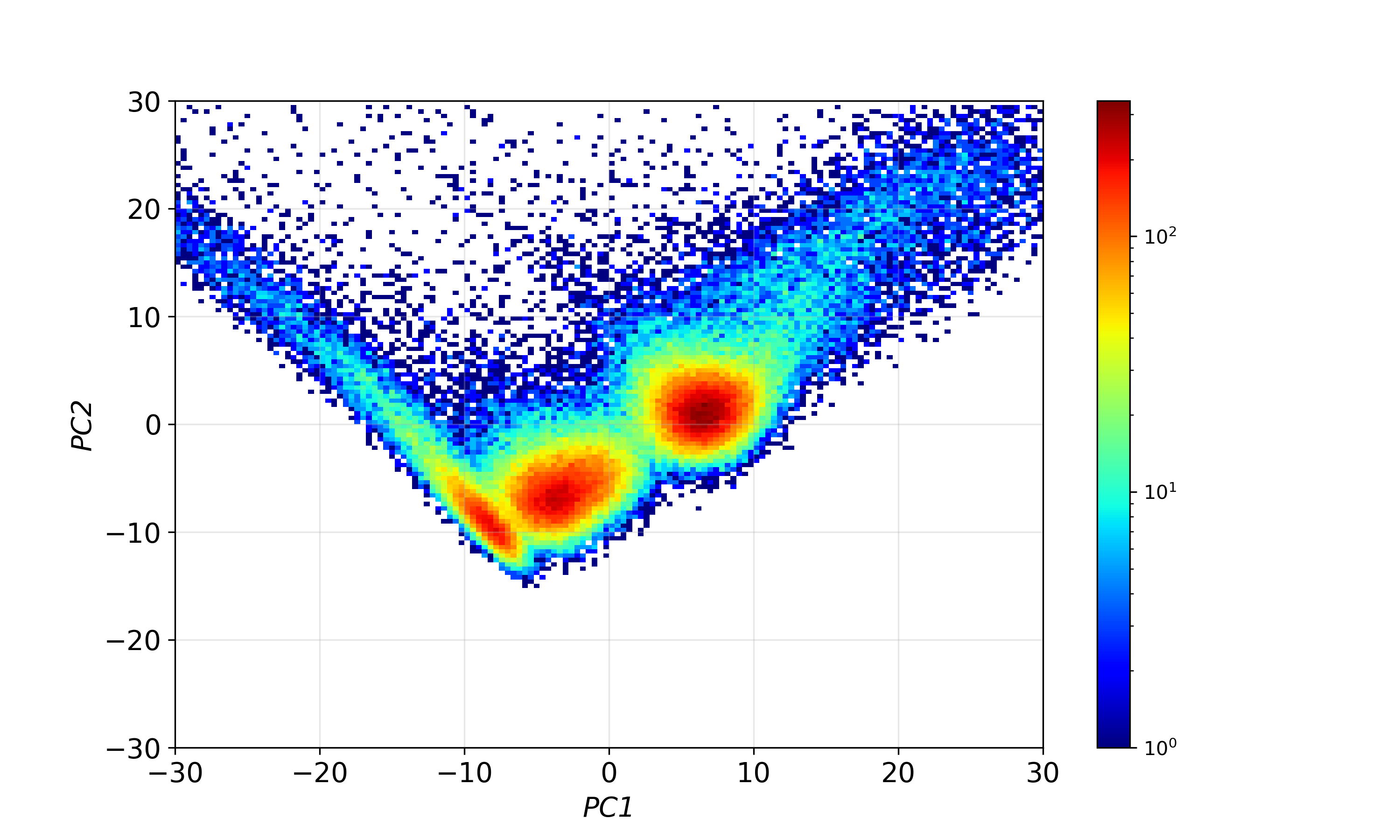}
        \caption{}
    \end{subfigure}%
    \hfill
    \begin{subfigure}[t]{0.55\textwidth}
        \centering
        \includegraphics[height=4.8cm, width=\linewidth]{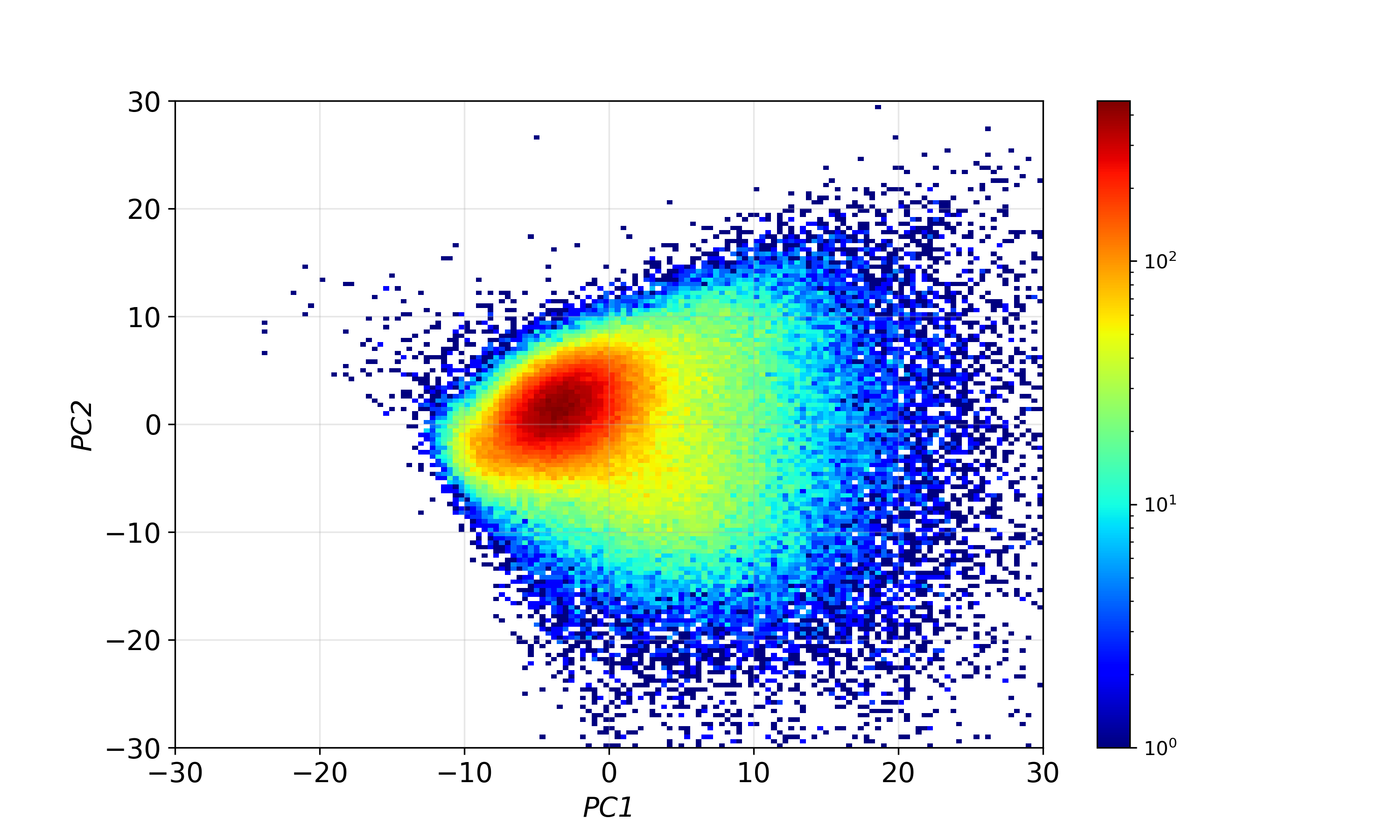}
        \caption{}
    \end{subfigure}

   \caption{Scatter plots of PC1 versus PC2 obtained from PCA analysis: (1) \textsuperscript{241}Am, where the dense clusters correspond to noise, the 26.5\,keV line, and the 59.5\,keV line (from left to right); (2) \textsuperscript{55}Fe, where the dense cluster corresponds to the 5.9\,keV line.}
  \label{fig:PCA}
\end{figure}

Each scintillation pulse was digitized at a sampling rate of \(10\,\text{MS/s}\), producing a waveform composed of discrete sample points. The total number of sample points defines the effective degrees of freedom for the analysis. For each dataset (run), PCA was applied to the ensemble of pulses using Python libraries, and the first two principal components (PC1 and PC2), corresponding to the largest eigenvalues, were used to construct a two-dimensional distribution. The resulting scatter plots, shown in Fig.~\ref{fig:PCA}, exhibit distinct dense clusters, or ``hot spots,'' which represent groups of similar pulses. These clusters correspond to characteristic emission lines of the calibration sources.  In Fig.~\ref{fig:PCA}. 1, three dense clusters are visible, with the leftmost corresponding to noise events. In contrast, a distinct noise cluster is not observed in the bottom panel, since the noise events are embedded within the main cluster associated with the 5.9\,keV emission line and are not dominant.

For the purpose of calibration and spectrum construction (Fig.~\ref{fig:spectrum}), we selected events from the densest regions of each dataset, thereby isolating signal-like pulses and rejecting outliers associated with noise. The selected pulses were then integrated over different time windows to evaluate light collection efficiency. As illustrated in Fig.~\ref{fig:template_Fit_Am}, an integration window of \(7\,\mu\text{s}\) at 175\,K was found to be optimal, providing a balance between maximizing scintillation light yield and minimizing contributions from electronic noise.

To evaluate the baseline resolution, the randomly triggered dataset was analyzed using the dataset which was taken in run 3. A Gaussian fit to the histogram of all sampled points yielded a standard deviation of \(\sigma \approx 1~\text{keV}\), corresponding to an effective energy threshold of approximately \(3\sigma \approx 3~\text{keV}\).

These results demonstrate the capability of the detector to reliably achieve an energy threshold as low as 3\,keV, highlighting its suitability for low-energy rare-event searches.

\section{Conclusion}

We have presented a detailed characterization of a \(1\,\text{cm}^3\) CsI(Tl) crystal coupled to a \(6\times6\,\text{mm}^2\) SiPM operated over a wide temperature range down to cryogenic conditions. The temperature-dependent study revealed a pronounced enhancement in the figure of merit near \(175\,\text{K}\), corresponding to the optimal balance between SiPM gain and CsI(Tl) light yield. Analysis of pulse templates confirmed that scintillation decay times increase at lower temperatures due to suppressed thermal untrapping of charge carriers from \(\mathrm{Tl}^{0}\) centers, consistent with carrier-recombination dynamics reported in prior studies.

The optimized setup at \(175\,\text{K}\) yielded a baseline resolution of \(1~\text{keV}\), corresponding to an effective energy threshold of \(3~\text{keV}\). These results confirm the system’s capability to operate with sub-few-keV thresholds, demonstrating its strong potential for low-energy rare-event searches, such as dark-matter or neutrino-interaction experiments, where low background and high sensitivity are essential.

Future work will focus on extending the measurements below \(175\,\text{K}\) by relocating the SiPM outside the cryogenic chamber and implementing optical-fiber coupling for light readout. This approach is expected to further improve baseline resolution by reducing thermal noise and electrical interference. In parallel, it enables us to developing a hybrid detector system capable of simultaneous phonon–photon readout.

\section{Acknowledgements}

The authors acknowledge the contributions of Maxim Koshelev and AtomSpectra for custom designing the SiPM amplifier circuit used in this study. We also gratefully acknowledge the financial support of the Mitchell Institute for Fundamental Physics and Astronomy.

\begingroup
\raggedright
\bibliographystyle{elsarticle-num}
\bibliography{ref}
\endgroup

\end{document}